**Hubble Telescope 30 Years in Orbit:  Personal Reflections**

Robert Williams

Space Telescope Science Institute  &  University of California/Santa Cruz     wms@stsci.edu

**Abstract:** With an initial requirement to make observations a minimum of 5-10 years, Hubble Space Telescope (HST) has continued to operate well for 30 years.  It has relied upon five servicing missions to repair and replace essential components.  Since the final Space Shuttle mission 10 years ago it has avoided major breaks in its operation, with the only serious effects of ageing in space being a progressive deterioration in the performance of the gyroscopes and sensitivity of the instrument detectors.  A number of factors were important in making HST a scientific landmark.  Ground-breaking discoveries have been made with HST----the most important being the discovery of cosmic acceleration.  When HST operation ceases future observations in space should be assured with successful operation of major missions now planned by NASA, ESA, and the China and Japanese Space Agencies.

## 1.  Uniqueness of Hubble Telescope

Space is a hostile environment for all objects----natural and man-made.  Damaging collisions between objects are inevitable.  Reactions between electronic equipment and the ionized upper atmosphere eventually lead to their failure.  UV radiation from the sun causes deterioration of all solid materials.  Temperature fluctuations for objects in low earth orbit also stress every type of components.  In spite of these obstacles, in April 2020 Hubble Space Telescope (HST) will have continued to operate remarkably well after thirty years in earth orbit.  The telescope did suffer from spherically aberrated optics for the initial three years following launch, and a significant part of its operational longevity was due to the fact it was serviced periodically by the NASA Space Shuttle.  Without the intervention of carefully trained astronauts renovating the telescope in orbit, HST would not have been a scientific success.  Fortunately, with servicing by the Shuttle the telescope was repaired and upgraded so it maintained its operation at original specifications.

 An important part of the international success of HST has been the official participation at 15% level of the European Space Agency (ESA) in the Hubble Project.  Given the international nature of astronomy truly major discoveries are enabled by the activities of scientists using different facilities.  In addition, scientists from ESA countries were able to participate in committees that set policies.  The international partnership of HST did lead to direct lines of communication being created between Space Telescope Science Institute (STScI) and most other major observatories and institutes so the great majority of the world's astronomers were kept knowledgeable of and felt a genuine part of the telescope.  Without official NASA partnership with ESA, HST would have had less impact in setting the directions of international astrophysics research for the past thirty years.

The location of HST above the atmosphere has been its great advantage.  The absence of atmospheric aberration of light gathered by HST results in its unmatched spatial resolution, in addition to the UV being accessible.  The background in the visible is significantly lower than even the best ground-based telescopes.  The development of adaptive optics on ground-based telescopes early in the life of HST did cause its high spatial resolution to become less a unique feature.  However, the practical limitation of small fields of view (FOVs) for adaptive optics on the largest ground-based telescopes did leave imaging over FOVs greater than approx 5-10 arcsec at spatial resolutions comparable to 0.1 arcsec the unique domain of the Hubble.  For extended objects this 'discovery territory' of HST has been directly responsible for many of the remarkable advances that occurred in many areas of astrophysics in the past generation.

## 2. Key Factors for Success

Looking back on the history of HST there have been a number of circumstances that combined, some by accident, to work together in a way that caused HST to become a scientific facility of historic significance. The principle reason for its success is obvious----as a telescope in space it avoids atmospheric distortions/aberrations to the light it receives. As some of the less obvious causes, I would cite:

a) The fact that the Space Shuttle program lacked a dedicated program or mission that justified its great expense. HST filled that void with far greater success than the International Space Station. After the historical Apollo program of landing humans on the moon, the Hubble servicing missions may stand out as the second greatest achievement of mankind in space.

b) Spherical aberration, which came so close to terminating HST, was saved by a truly remarkable collaboration between NASA, STScI, the astronomical community, and the contractors who built the telescope and instruments. A number of scientists, engineers, managers, and astronauts were suddenly thrust into roles in which they performed superbly under huge political and public pressure. The success of the first servicing mission, with all the media attention it received by the public, brought the value of HST as a premier facility to all people. In hindsight, the transformation of the failed telescope into its unique discovery mode played a large role in providing healthy funding levels for the Hubble Project until the present time.

c) The creation of an independent institute, STScI, to manage the scientific program of the Hubble. NASA had exercised management of previous astronomical missions, such as IUE and the OAOs, and did a very credible job of getting good science out of their missions. But, HST was almost two orders of magnitude more complex and expensive than previous NASA missions. There was concern that NASA, which is mission oriented and therefore frequently compromised when operating existing facilities, might turn its attention away from HST in order to start new projects and missions. The U.S. astronomical community wanted to have independent control over the HST science program and also the ability to create a unified national lobby for funding with Congress, which NASA is forbidden to do on its own behalf. The community has very ably provided consistently strong support for every aspect of the Hubble Program.

d) A large outreach and education program that focused on both the media and the internet. After the first servicing mission corrected HST's optical problem the opportunity to make spectacular colored images of celestial objects available to the public became obvious. The steady flow of remarkable images captured the public world-wide, and especially young students. It would have been difficult for even the proverbial hermit living in a cave to escape the amazing universe that has been revealed by the Hubble and collaborating telescopes!

e) The availability of Director's Discretionary (DD) time on HST. By contract with NASA the STScI Director is authorized to schedule up to 10% of the observing time. This allows risky and long-term projects to be carried out that would otherwise be rejected by telescope allocation panels, who have a tendency to try to schedule as many observing proposals as possible, thereby discriminating against certain types of proposals. Both the Hubble Deep Field and the observations of SNe Ia that demonstrated cosmic acceleration were initially undertaken with DD time.

f) Large data reduction funding provided by NASA for analysis of all HST data for U.S. based researchers. An annual budget of order U.S. $40 million has been made available to any researcher based at a U.S. institution in support of their work on an approved Hubble program. This funding

has had a dramatic impact in providing the resources needed to convert Hubble data into useful scientific results.

The above examples are just some of the factors that have produced HST's high scientific productivity. The great legacy of HST's 30 years of observations are the major discoveries that have changed our perception of the universe. The most important of these is widely considered to be the discovery of cosmic acceleration, driven by a not yet fully understood source of energy. That result did not come about simply. It relied on the diligent work of two large teams having a broad range of expertise. Having had a role in making decisions that determined how HST would be used in the campaigns to make the discovery of Dark Energy (Williams, Okamura, & Suzuki 2016), I would like to take the opportunity on this 30th anniversary of the Hubble's operational life to recount certain facts for the historical record about how those observational programs came about.

## 3. The Accelerating Universe

One of the early Key Projects on HST was led by PIs W. Freedman and J. Mould before the first servicing mission that corrected Hubble's spherical aberration. The goal of that project was to determine the Hubble constant to an accuracy of 10%. On the success of the servicing mission, two independent teams were formed that proposed to extend the work of the Freedman/Mould Key Project by looking for changes in the cosmic expansion rate. Focusing on the deceleration parameter, $q_o$, the teams set out to make photometric observations of distant Type Ia supernovae (SNe Ia), using them as standard candles to determine their distances, i.e., lookback times, followed by acquisition of their spectra to determine their radial velocities.

At a time when cosmic acceleration was not yet suspected, early analysis of observations of the initial set of high redshift, z>0.35, SNe Ia by the Supernova Cosmology Project (SCP) team, which had all been acquired on ground-based telescopes, did not demonstrate any evidence for a change in $H_o$ with time (Perlmutter, Gabi, Goldhaber et al. 1997). The photometric uncertainties from the ground-based telescopes were too large. The SCP group, led by Saul Perlmutter and with more than 30 members, had already realized from their early analyses that HST with its higher spatial resolution and lower background light would provide more accurate brightness levels that could better reveal any change in the cosmic expansion rate. For this reason, Perlmutter and the SCP team proposed for 23 orbits of HST observations to study distant supernovae to the HST Cycle 6 Telescope Allocation Committee (TAC) in November 1995. Their proposal was made possible by the unique process Perlmutter and colleagues had pioneered in detecting distant SNe at the time of their outbursts with ground-based telescopes (Perlmutter et al. 1994).

A year earlier another team interested in supernovae had been formed, called SINS (Supernovae INtensive Studies) and led by Robert Kirshner. Their early interest was in making HST observations in the UV to study both Type I and II SNe to better understand how the hydrodynamics of the outburst affects the post-outburst radiation field (Kirshner et al. 1993; Jeffery et al. 1994). In 1993, one of the SINS team members, Mark Phillips at Cerro Tololo Observatory in Chile, had published what has become a seminal paper defining the relation between the peak luminosity of SNe Ia and their brightness 15 days after maximum brightness (Phillips 1993). The importance of Phillips' work in refining SNe Ia as standard candles led members of the SINS team to begin shifting their primary interest to determining the value of the Hubble constant over time, as the SCP team was attempting, as much as understanding the physics of SNe outbursts. A number of members of the SINS team eventually formed an additional team that would become the 'Hi-z' team, led by Brian Schmidt and Adam Riess with members M. Phillips, N. Suntzeff, R. Kirshner, A. Filippenko, among others. The SINS-to-Hi-z transformation was taking place in 1995 as the SINS team proposed to the HST Cycle 6 TAC in November 1995 for 30 orbits of primarily ultraviolet observations to study SNe I and II, including SNe 1987A, 1992A, and 1993J. HST Cycle 6 observations were to be taken in the period 1 July 1996 - 30 June 1997.

The Cosmology panel of the Cycle 6 HST TAC evaluated and graded both the SINS proposal and the SCP proposal, giving good marks to each proposal. The ranking of the SINS proposal was sufficiently high that it was approved for Cycle 6 observations. The ranking of the SCP proposal was lower than that of the SINS proposal and below the cut-off line for approval in that Cycle due to the oversubscription factor for HST observations, which was very high: (orbits requested)/(orbits available) $\cong 7$. Thus, the Perlmutter team proposal did not receive Hubble time for that Cycle. The two teams were notified of the TAC recommendations, to which I as Director gave final approval, in December 1995.

In January 1996 Saul Perlmutter approached me at the San Antonio meeting of the American Astronomical Society to discuss a proposal he wished to submit for DD time related to his work using distant SNe Ia to determine the distance scale and the deceleration parameter. Saul explained to me his proposal with conviction, believing that HST was a key to determining how $H_o$ might be changing in time. He did acknowledge that his new proposal was similar to and improved from the SCP proposal that had not been successful in Cycle 6. At the end of our conversation I invited Saul to submit his proposal for DD time in spite of the informal policy we had instituted at STScI that we not normally consider DD time for proposals that had not been successful in the recent TAC process. Saul did submit the proposal to the Institute in early February, and I acknowledged its receipt.

Following each annual TAC meeting there are normally a large fraction of the observing proposals not approved because of the huge oversubscription of available HST time. It became normal procedure for a number of these unapproved proposals to be immediately submitted to the Director for consideration for DD time. Saul's was one such proposal. During the period following correction of the spherical aberration, I preferred to save DD time for new initiatives and time-critical observations. After reading the Perlmutter SCP DD proposal, together with the critical TAC review comments on their Cycle 6 proposal, my decision was to not approve his DD proposal at that time.

The situation changed several months later in May 1996 when the annual STScI May Symposium took place and caused me and many others to become more excited about more extensive observations of SNe with HST. The symposium was devoted to the topic 'The Extragalactic Distance Scale' and there were excellent talks given on improved values of $H_o$ and the likelihood that HST observations could reveal the deceleration parameter, providing a determination of the mean density of the universe. A great deal of enthusiasm was generated at that symposium for HST as a unique tool to be used for cosmological studies (Livio, Donahue, & Panagia 1997). I must admit to having been too short-sighted to appreciate Saul's foresight when we discussed this topic in San Antonio. By the end of the symposium I had become convinced that HST should indeed devote a significant effort to determining not just $H_o$ better, as the Freedman and Mould Key Project was doing, but also $q_o$----which is what Saul had been advocating in his rejected Cycle 6 and DD proposals.

It is worth noting that this situation occurred shortly after the large Hubble Deep Field (HDF) program, necessarily undertaken with a large amount of DD time, had just been completed and made public. Thus, my initial instinct was that the difficult challenge of measuring a change in the cosmic expansion rate was likely a similar situation to the HDF, i.e., best attacked with HST using a substantial allotment of DD time. Waiting until the next Cycle 7 TAC process would mean that no observations would be made for the next year and a half. I therefore made the decision near the end of the May Symposium to gather together some of the attendees for a discussion about using HST DD time to jump start serious initiatives to measure what at that time was believed to be cosmic deceleration.

Three people who during the week had spoken to me most energetically about using the Hubble for cosmic expansion observations were invited to my office to talk about ways forward. They were my close colleague at the Institute, Nino Panagia, a member of Perlmutter's SCP team, and my former CTIO colleagues Mark Phillips and Nick Suntzeff, members of the SINS team and part of the group who were already in the process of forming the Hi-z team. Bob Kirshner, who spoke at the symposium and was the SINS team leader, was also invited. Not in attendance at the symposium were Saul Perlmutter, Adam Riess, Brian Schmidt, Allan Sandage and others, who had they been in Baltimore would also have been invited to join our discussion. My notes of the meeting indicate that when I raised the question of allocating DD time to the problem three of the participants were enthusiastic about the capabilities of the Hubble in observing distant SNe Ia to determine variations in $H_o$. One person was not: Bob Kirshner. Bob believed that HST was not needed. If I can paraphrase his comments, they would be something like, "HST is a valuable resource being used to solve many important problems in astrophysics. Ground-based photometry of distant SNe should be sufficiently accurate to address the question of $q_o$. No need to divert the valuable resource that is HST to this problem." [1]

Kirshner's confidence in ground-based photometry was not without basis. The SCP group began their quest to observe cosmic velocity changes using telescopes on the ground---partly because of the Hubble's spherical aberration. Their initial inconclusive results did not become apparent and were not disseminated to the community until they had finalized their analysis and submitted their paper in 1996 (Perlmutter et al. 1997). This was some months after the discussion in my office in May.

At the conclusion of the discussion in my office at the end of the May Symposium I was convinced that an allocation of HST DD time was likely to make an important contribution to determining variations in $H_o$. I must admit that the recent success of the Hubble Deep Field, which was carried out entirely with 150 orbits of DD time, was influencing my thinking (Williams 2018). DD orbits are for risky observing programs that are unlikely to move forward by TAC approval in the very competitive HST environment. Too many orbits would be needed, and requests for large allotments of times are disadvantaged when approval requires consensus within a large committee. I closed our office discussion by offering 28 orbits of DD time to both the SCP and Hi-z team if they could submit to me well-reasoned proposals for that amount of orbits to begin attacking the problem of changes in $H_o$.

To make a long story short, both SCP and Hi-z teams did submit proposals that were received in late summer of 1996. I reviewed them; was positively impressed with each; and awarded 28 orbits of Cycle 6 DD time for both programs to be carried out. The scheduling of the HST observations was complicated because it needed to be coordinated with previous ground-based observations where the SNe were discovered. The first observations were delayed by the second HST servicing mission in February 1997, and took place in late May of that year

Analysis of the early observations were quite promising such that both the SCP and Hi-z teams received generous allocations of HST observing time via the normal GO TAC process in following years. And for the historical record it is of some interest to note that prior to the submission of the original Hi-z proposal for the DD orbits I was offering both teams, I received an email from Dr. Brian Schmidt, PI of their proposal, on 6 June 1996 in which he wrote: "On behalf of the High-Z SN Search Team, I would once again like to thank you for encouraging us to apply for Director's Discretionary time to follow distant Type Ia supernovae. We believe that a 28 orbit program …. can serve as a stepping stone to a larger HST program to map out the deceleration of the Universe so that we can get a handle on not only $q_o$, but Lambda as well." Of course, both teams did outstanding work in fulfilling the above belief in success (Riess et al. 1998: Perlmutter et al. 1999), to which a number of highly regarded international prizes attest.  But it is notable that in the middle of 1996 the teams did not think of acceleration at all----it was deceleration they were after!

---

[1] After further fact finding and discussion with colleagues in the ensuing weeks, Dr. Kirshner completely changed his thinking and became an advocate of using HST to find changes in the Hubble constant.

## 4.  The Future

HST is operating well on its 30th birthday, with various components such as the gyroscopes and detectors showing deterioration with increasing age.  Work arounds have been found that allow observations to be made that compensate for the decreased sensitivity of the instruments and ability of the telescope to hold its steady lock on targets.  The number of publications and citations of papers published using Hubble data continue to increase, with the fraction of papers published using results based on data from the HST MAST Archive exceeding the number based on direct observations.  Whenever HST finally shuts down operations due to some major failure its data archive will constitute a monument to astronomy that will be mined for decades----much like the iconic Palomar Sky Survey of the 1950s.

With the retirement of the Space Shuttle there are no future manned missions possible for HST.  The current suite of instruments and electronic hardware will operate until they fail as long as funds are made available by NASA to continue the operation of the telescope.  The final servicing mission in 2009 did install a ring structure on the back end of the telescope that can be used to attach a robotic de-orbiting propulsion module whose ignition will produce a programmed deceleration of HST, causing it to fall harmlessly into the ocean.

There is great value in having both the Hubble and James Webb space telescopes operate simultaneously on the same objects and fields.  Working together, HST's instruments in the visible and JWST's infrared instruments would provide data that enable much stronger analyses due to improved diagnostics from the broader wavelength coverage and the aperture masking interferometer capability on JWST.  With good fortune HST should continue to operate into the JWST era so the two telescopes can work together to unmask the cosmos.  When HST operations cease future observations in space will be assured with successful operation of major missions now planned by NASA, ESA, and the China and Japanese Space Agencies.

**Figures**

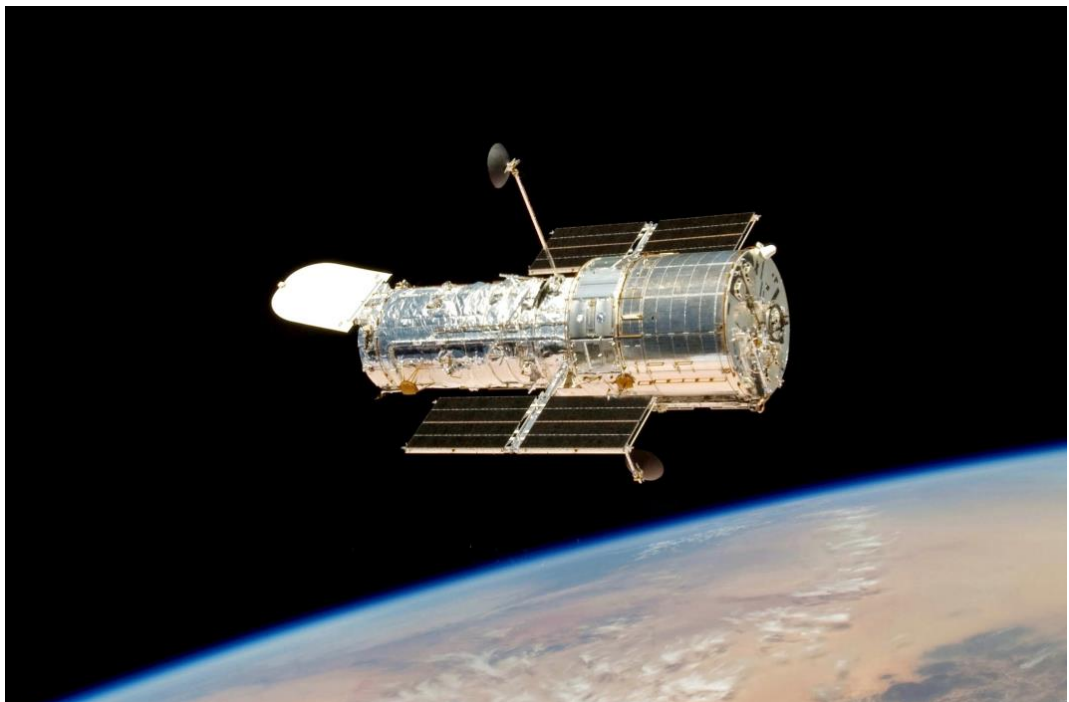
Fig. 1 - Hubble Telescope at the end of the final servicing mission, May 2009.

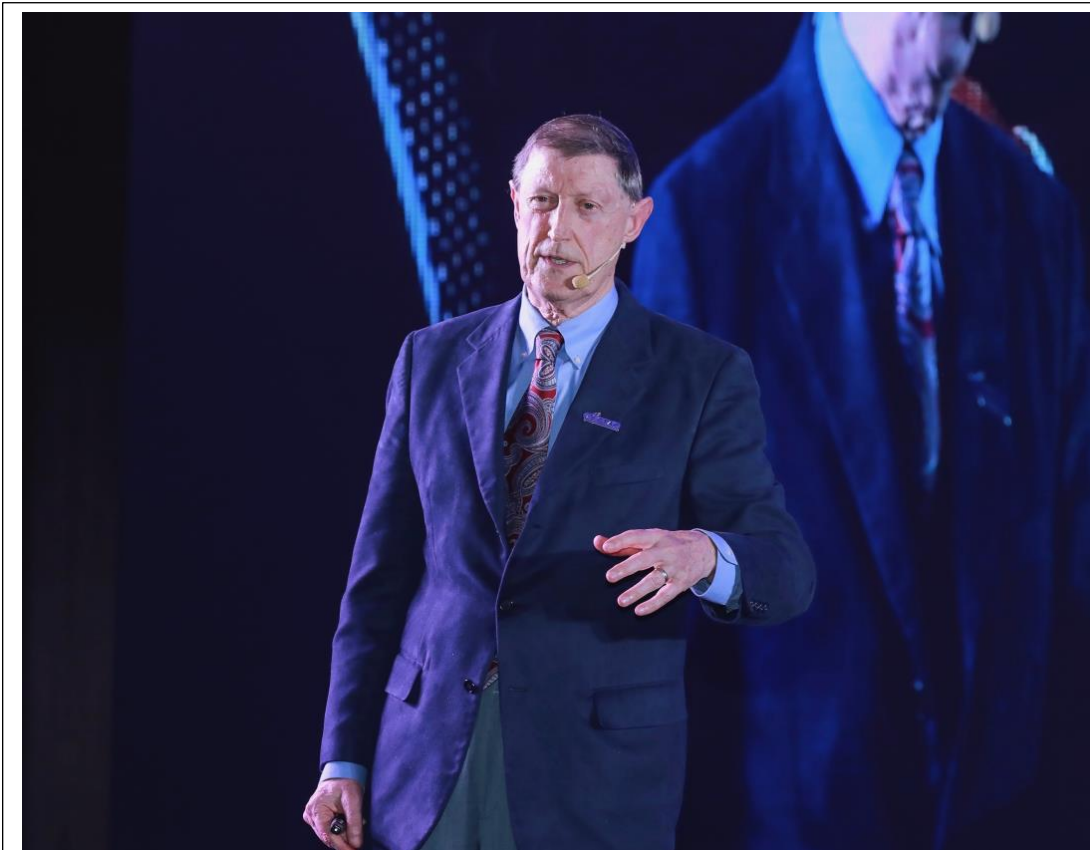

Fig. 2 – Author Robert Williams lecturing on Hubble Space Telescope science in China in 2019.

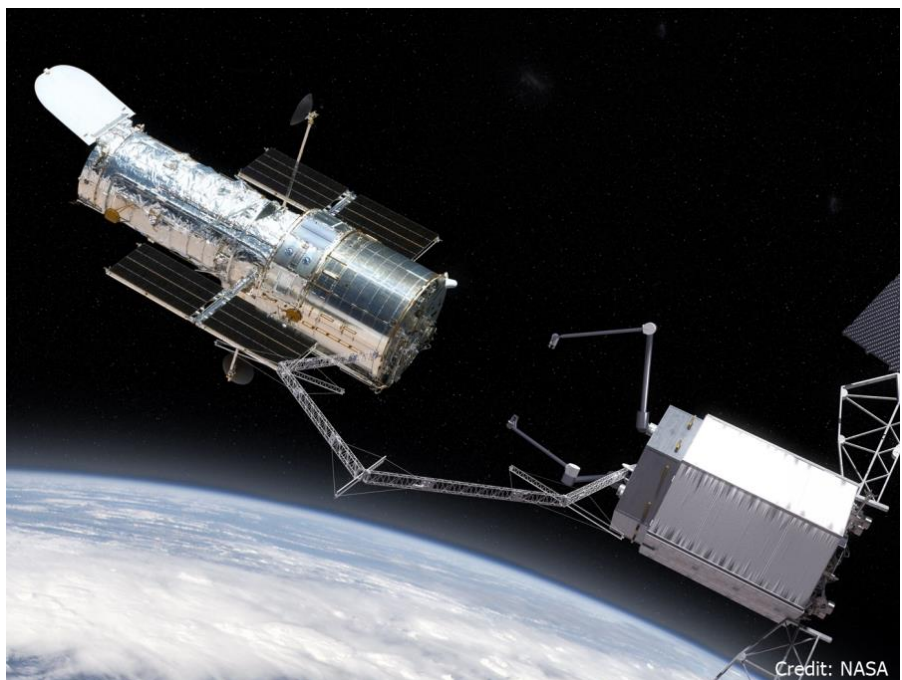

Fig. 3 – HST has a coupling ring that will accept a robotic propulsion module that can de-orbit the telescope when it becomes non-operational.